\documentclass[11pt,twoside]{article}
\usepackage{asp2010}

\resetcounters

\begin{document}

\title{Creation of neutral disks during outbursts of symbiotic binaries}
\author{Zuzana~Carikov\'{a} and Augustin~Skopal
\affil{Astronomical Institute, Slovak Academy of Sciences, 059 60 Tatransk\'{a} Lomnica, Slovakia}}

\begin{abstract}
Multiwavelength modelling of the spectral energy distribution
of symbiotic binaries suggests that a neutral disk-like
zone is created around the hot star near the orbital plane
during active phases.
Presumably, this is connected with the enhanced wind from the
active hot star.
To test this idea, we applied the wind compression model to
active hot stars in symbiotic binaries, within which
the wind particles are compressed more to the equatorial plane
due to a fast rotation of the hot star.
Accordingly, we calculated the ionization structure for such
compressed wind and ionizing photons from the hot star.
We found that the hot star wind, enhanced during active phases
to $\sim (10^{-7} - 10^{-6})$ ${\rm M}_{\sun} {\rm yr}^{-1}$, and the rotational
velocity of 100 -- 350 km s$^{-1}$ at the star's equator
lead to formation of a neutral disk-shaped zone.
The presence of such disks is transient, being connected
with the active phases of symbiotic binaries.
During quiescent phases, such neutral disks cannot be
created, because of insufficient mass loss rate.

\end{abstract}

\section{Introduction}

Symbiotic stars are long-period interacting binary systems,
which comprise a late-type giant and a hot compact star, which
is in most cases a white dwarf (WD). Accretion from the giant's
wind makes the WD to be a strong source of ionizing radiation.
During so-called quiescent phases, it ionizes a fraction of the
giant's wind giving rise to a dense nebula extended within the
binary \citep{stb}. The temperature of the hot star is
$T_{\rm h} \sim$ 10$^{5}$ K during quiescence.
However, during active phases, the ionization structure
changes drastically. In symbiotic binaries with high orbital
inclination we observe a two-temperature-type of the UV
spectrum. The cooler component is produced by a relatively warm
stellar source ($T_{\rm h} \sim$ 22 000 K), while the hotter one
is represented by the highly ionized emission lines and a strong
nebular continuum.
This situation suggests the presence of a neutral disk-like
structured material surrounding the accretor and hot emitting
regions located above/below the disk, which is seen approximately
edge-on \citep[see Fig. 27 of][]{sk05}. A rapid creation of such
ionization structure during the first days/weeks of outbursts is
connected with the enhanced hot star wind. During quiescence the
mass loss rate from the hot star is
$10 ^{-8}$ M$_{\sun}$ yr$^{-1}$, but in activity it increases to
$\sim (10^{-7} - 10^{-6})$ M$_{\sun}$ yr$^{-1}$ \citep{sk06}.
The wind gives rise to the optical bursts by reprocessing
high-energy photons from the Lyman continuum to the optical/UV
\citep{sk+09}.

The rotation of the hot star with radiation driven wind leads to
compression of the outflowing material towards equatorial
regions due to conservation of the angular momentum.
This wind compression model was developed by \citet{bjorkcass93}.
If the streamlines of gas from both hemispheres do not cross the
equator then we are talking about the wind compressed zone (WCZ)
model described by \citet{igncassbjork96}.
In this contribution, we test the idea if the compression of the
enhanced wind from the rotating hot star towards the equatorial
regions could create such neutral disk-like structure around the
hot star in symbiotic binaries during their active phases, and
thus to simulate the above-mentioned hot object structure based
on modelling SED.

\section{Density in the wind compression model}

Density in the hot star wind as a function of polar coordinates
($r$, $\theta$) follows from mass continuity equation
\begin{equation}
  N_{\rm H}(r,\theta) = 
          \frac{\dot{M}}{4\pi r^2 \mu_{\rm m} m_{\rm H}
          v_{\rm r}(r)}\left(\frac{d\mu}{d\mu_{0}}\right)^{-1},
\end{equation}
where the compression of the wind is described by the geometrical
factor $d\mu/d\mu_{0}$ which depends on the parameters of the
wind as well as on the rotational velocity of the hot star,
$\dot{M}$ is the mass loss rate of the active hot star,
$\mu$ is the mean molecular weight and $m_{\rm H}$ is the mass
of the hydrogen atom.
For the radial component of the wind velocity $v_{\rm r}(r)$
we adopted $\beta$-law
\begin{equation}
  v_{\rm r}(r) = v_{\infty}\left(1-\frac{bR_{\ast}}{r}
                           \right)^{\beta}, 
\end{equation}
where
\begin{equation}
  b = 1-\left(\frac{a}{v_{\infty}}\right)^{\frac{1}{\beta}},
\end{equation}
where $a$, $v_{\infty}$ are the initial and terminal velocity
of the wind, respectively and $R_{\ast}$ is the radius of the
hot star. More details about the wind compression model can be
found in \citet{lamcass99}.

\section{Ionization structure in the wind}

Here we investigate the case, when the hot star is the source
of both the ionizing photons and the wind particles.
The ionization boundary is defined by the locus of points at
which ionizing photons are completely consumed along path
outward from the ionizing star.
We calculated the ionization boundary
by the equation of photoionization equilibrium which equals
the number of ionizations with the number of recombinations.
We assumed optically thick case (so called "Case B" or
"on-the-spot approximation"). For the sake of simplicity, we
assumed that wind contains only hydrogen atoms.
Further we define the distance from the center of the hot star
in units of its radius $R_{\ast}$,
i.e. $r \rightarrow u=r/R_{\ast}$.
For calculating ionization boundary in the hot star wind we
derived an equation, which can formally be written as
\begin{equation}
  X = f(u,\theta,a,v_{\infty},\beta,v_{\rm rot}),
\label{eqn:calcX}  
\end{equation}
where $u$, $\theta$ are polar coordinates ($\theta=0$ at the
rotational axis), $a$, $v_{\infty}$, $\beta$ are parameters
of the wind, $v_{\rm rot}$ is the rotational velocity of
the hot star, and the value of the parameter $X$ is given by
\begin{equation}
  X = \frac{8\pi \mu_{\rm m}^{2} m_{\rm H}^{2}}{\alpha_{\rm B}}
  R_{\ast} L_{\rm H} \left(\frac{v_{\infty}}{\dot{M}}\right)^{2},
\label{eqn:X}
\end{equation}
where $L_{\rm H}$ is the rate of photons from the hot star, 
capable of ionizing hydrogen (it is given by the temperature
$T_{\rm h}$ and luminosity $L_{\rm h}$ of the ionizing source)
and $\alpha_{\rm B}$ is the total hydrogenic recombination
coefficient in case B.
To achieve complete ionization boundary in the hot star wind
one has to solve Eq. (\ref{eqn:calcX}) for each direction
$\theta$ separately.

Most of the parameters of the active hot star and its wind in
Eqs. (\ref{eqn:calcX}) and (\ref{eqn:X})
can be determined from observations. However the rotational
velocity of the hot star in symbiotic binaries,
required for Eq. (\ref{eqn:calcX}), is not commonly known.
Lower limit is given by the rotational velocities of the
isolated white dwarfs and upper limit is given by the condition
of the considered WCZ model, i.e. streamlines of gas do not
cross the equator.
We found to be the most appropriate to calculate models for
different rotational velocities from 100 to 350 km s$^{-1}$.

In Fig. 1 we show some examples of ionization boundaries for
different values of the parameter $X$ resulting in creation
of a neutral disk-like zone. We illustrate this case for two
different rotational velocities of the central hot star,
200 and 300 km s$^{-1}$, respectively.

\begin{figure*}
\begin{center}
\begin{tabular}{cc}
\resizebox{6.0cm}{!}{\includegraphics[angle=270,trim=0.3cm 3.5cm 0.3cm 4cm]{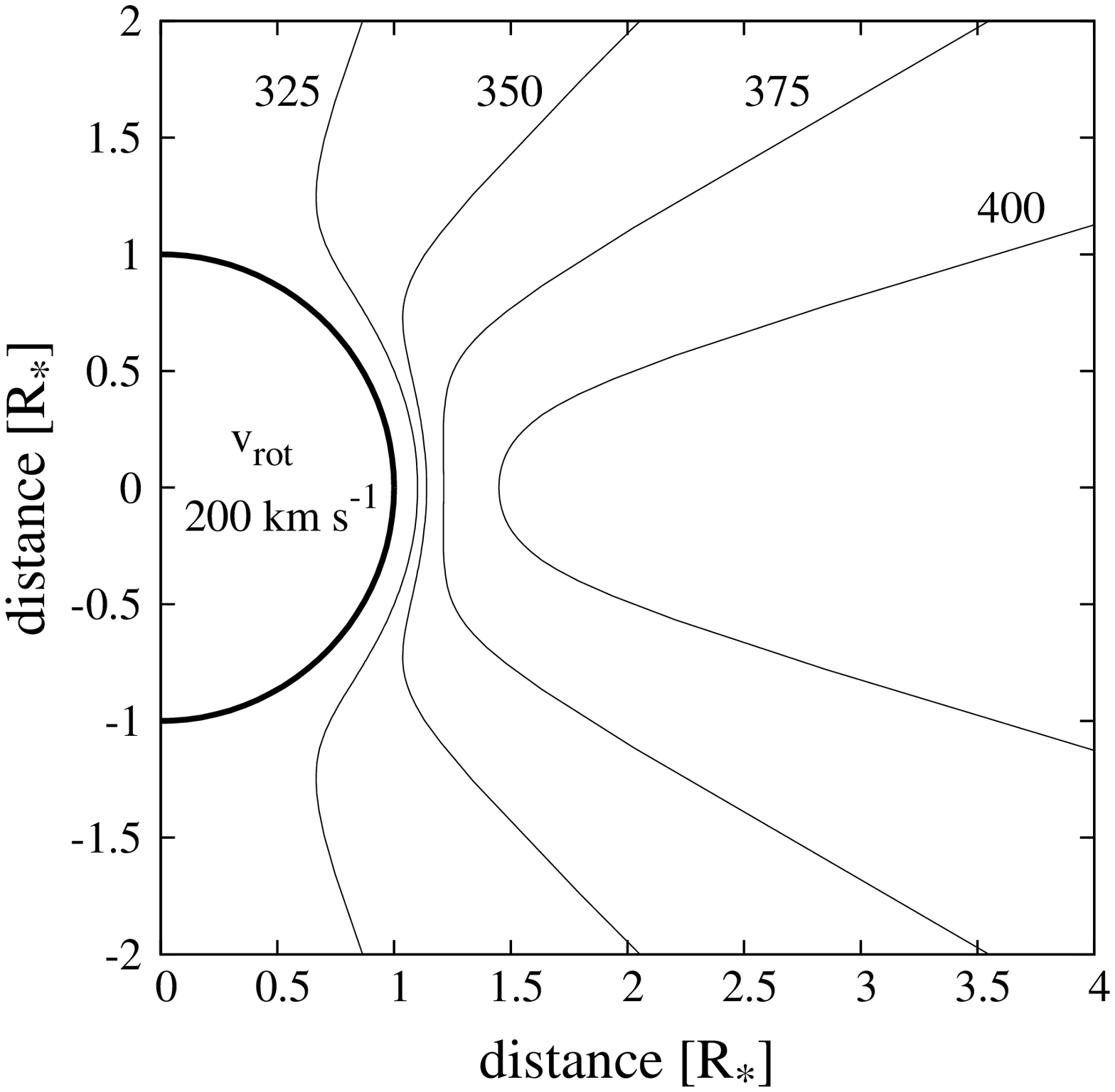}}&
\resizebox{6.0cm}{!}{\includegraphics[angle=270,trim=0.3cm 3.5cm 0.3cm 4cm]{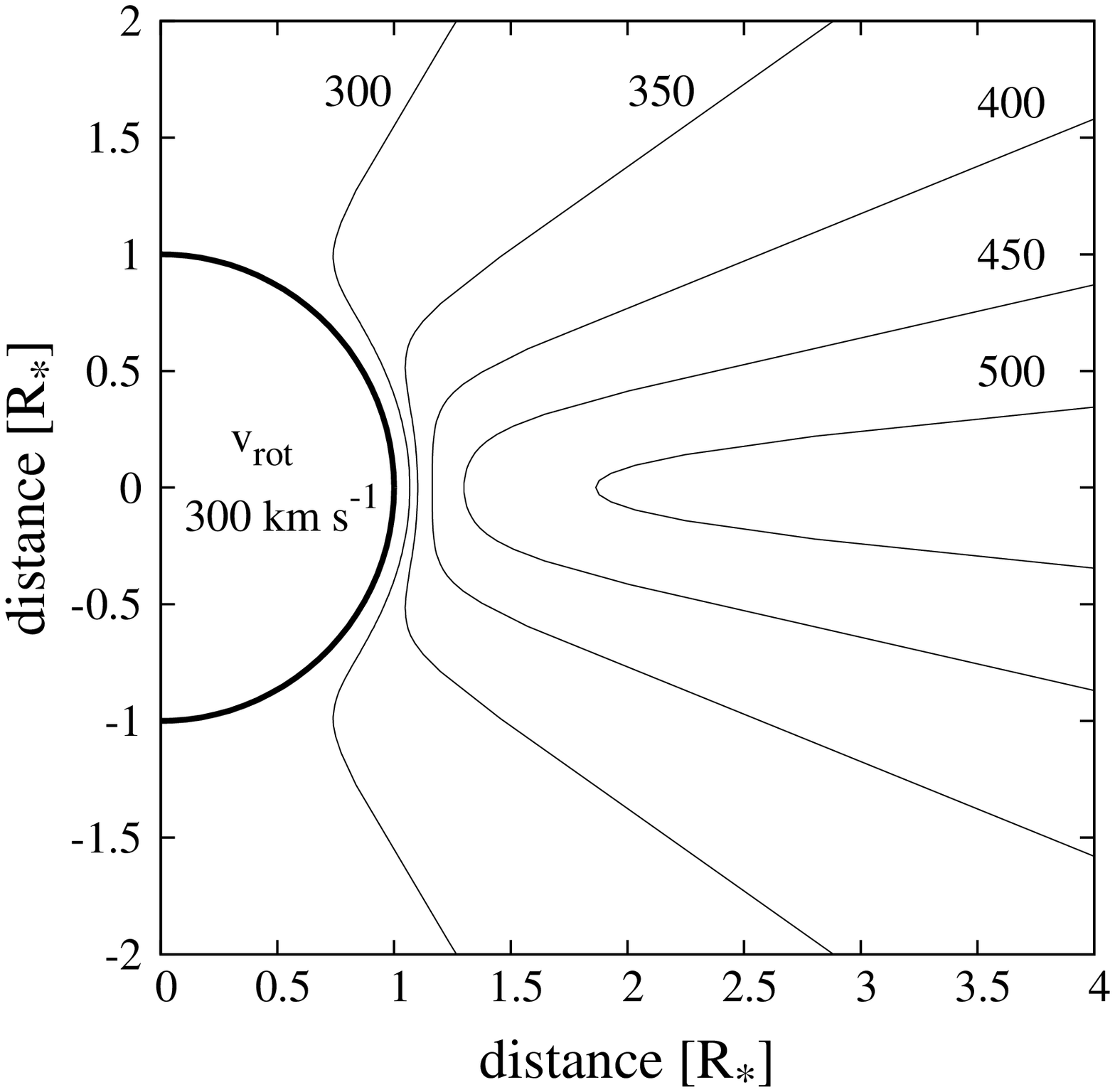}}
\end{tabular}
\end{center}
\caption[]{Examples of ionization boundaries in the wind for two
different rotational velocities of the central hot star.
Left: $v_{\rm rot}=$ 200 km s$^{-1}$.
Right: $v_{\rm rot}=$ 300 km s$^{-1}$.
Individual ionization boundaries are labeled by the value of
the parameter $X$. From an ionization boundary towards the pole
of the star (y-axis) there is the ionized zone, and towards the
equatorial plane (y $=$ 0) there is the neutral zone.
Ionization structures are axially symmetric with respect to the
polar (rotational) axis of the hot star.
Distances are in units of radius of the active hot star $R_{\ast}$.}
\label{fig:exam}
\end{figure*}

\section{Discussion}

We revealed that there is only a certain range of values of the
parameter $X$, for which the neutral disk-like structure can be
created near to the equatorial plane of the active hot star.
For higher rotational velocities of the hot star (i.e. higher
compression of the hot star wind towards the equatorial plane)
this range is wider.
For a given rotational velocity, small values of the parameter
$X$ ($\sim 200$) correspond to ionization boundaries, which are
enclosed at a vicinity of the hot star. On the other hand,
increasing value of $X$ corresponds to moving the ionization boundary
away from the vicinity of the active hot star, as well as to
decreasing the opening angle of the neutral disk-like zone
until it disappears for very high values of $X$.
However, particular values depend also on the parameters of
the active hot star wind.

\begin{figure}
\begin{center}
\resizebox{10.0cm}{!}{\includegraphics[angle=270]{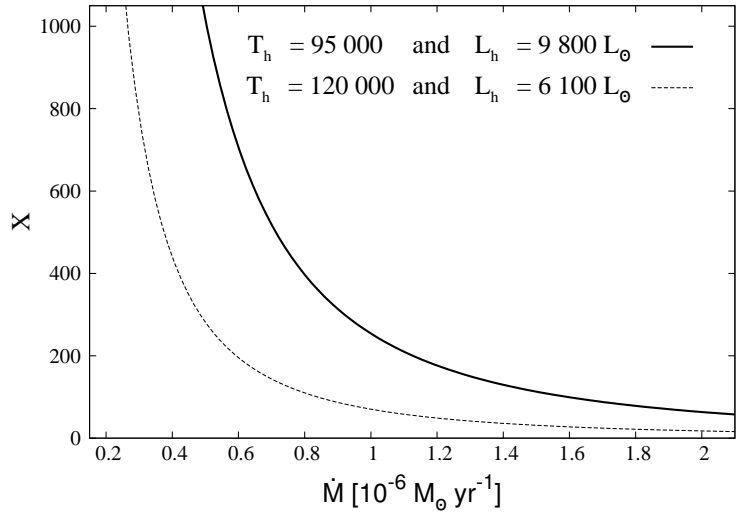}}
\end{center}
\caption[]{The value of the parameter $X$ as a function of
the mass loss rate from the hot star $\dot{M}$ for two different
combinations of the temperature $T_{\rm h}$ and luminosity
$L_{\rm h}$ of the active hot star as the ionizing source.}
\label{fig:xmdot}
\end{figure}

Our calculations (e.g. Fig. \ref{fig:exam}) showed that creation
of the neutral disk-like structure requires the value of the
parameter $X$ to be of the order of hundreds.
Figure \ref{fig:xmdot} shows the dependence of the parameter
$X$ on the mass loss rate from the hot star, $\dot{M}$.
We calculated these dependences for observed parameters of the
hot star and its wind, where we used two different combinations
of its temperature $T_{\rm h}$ and luminosity $L_{\rm h}$, which
were taken from \citet{sok06}.
From Fig. \ref{fig:xmdot} we can see that observed mass loss
rates from the active hot star,
i.e. $\sim (10^{-7} - 10^{-6})$ ${\rm M}_{\sun} {\rm yr}^{-1}$,
correspond to the required values of the parameter $X$.

According to Eq. (\ref{eqn:X}), the parameter $X$ strongly
depends on the mass loss rate of the hot star $\dot{M}$ as
\begin{equation}
  X \propto \frac{1}{\dot{M}^{2}}.
\end{equation}
The mass loss rate of the hot star during quiescent phases is
by 1 - 2 orders lower than during active phases, which enlarges
the parameter $X$ so high that no neutral disk-like structure
can be created during quiescence.

\begin{figure}
\begin{center}
\resizebox{10.0cm}{!}{\includegraphics[angle=270]{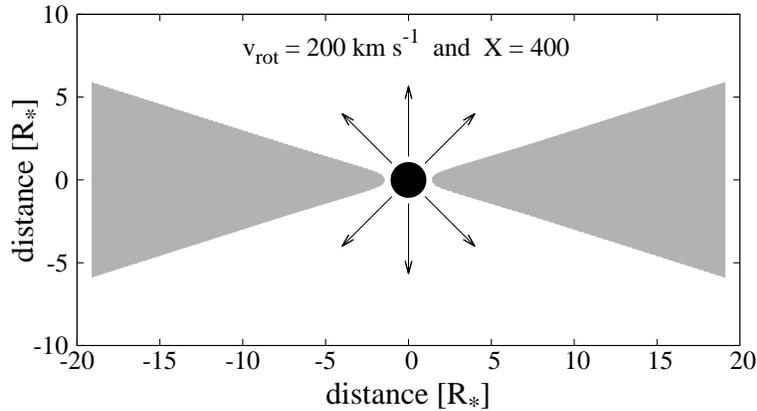}}
\end{center}
\caption[]{An example of ionization structure around the active
hot star calculated for $v_{\rm rot}=200$ km s$^{-1}$ and $X=400$
($\dot{M} \sim 6\times10^{-7}$ ${\rm M}_{\sun} {\rm yr}^{-1}$),
which is similar to the schematic one proposed on the basis of
multiwavelength modelling of the spectral energy distribution
of symbiotic stars \citep[see Fig. 27 in][]{sk05}.
White dwarf is denoted by the black filled circle and
modelled neutral disk-like structure is grey. Material in the
regions above and below this structure is ionized. Distances
are in units of the radius of the central hot star $R_{\ast}$.}
\label{fig:col}
\end{figure}

\section{Conclusion}

On the basis of the multiwavelength modelling of the spectral
energy distribution of symbiotic binaries, \citet{sk05} suggested
that neutral disk-like zone is created around the hot star near
the orbital plane during active phases. Modelling the broad
H$\alpha$ wings from the optically thin wind of the hot star
showed enhancement of its stellar wind  by a factor of $\geq 10$
in activity \citep{sk06}.
We found that the compression of the enhanced stellar wind
in active phases from the rotating hot star towards
equatorial regions can lead to the creation of the neutral
disk-like structure around the active hot star.
Figure \ref{fig:col} shows an example of ionization structure
around the active hot star in symbiotic binaries calculated for
$v_{\rm rot}=200$ km s$^{-1}$ and $X=400$
($\dot{M} \sim 6\times10^{-7}$ ${\rm M}_{\sun} {\rm yr}^{-1}$).
Its shape is very similar to that shown in the schematic
figure proposed by \citet{sk05}.
Finally, we explained that due to a low mass loss rate from
the hot star, no neutral disk-like structure can be
created during quiescent phases.

\acknowledgements This research was supported by a grant of the
Slovak Academy of Sciences, VEGA No. 2/0038/10.

{}
\end{document}